\documentclass{sig-alternate}
\usepackage{comment}
\usepackage{subfigure}

\begin{document}

\title{Being Rational or Aggressive? A Revisit to Dunbar's Number in Online Social Networks}
%
%
%
%
%
\numberofauthors{5} 
%
\author{
%
%
\alignauthor
Jichang Zhao\\
       \affaddr{Beihang University}\\
       \email{zhaojichang@nlsde.buaa.edu.cn}
\alignauthor
       \setlength{\parindent}{2em}
       Junjie Wu\\
       \setlength{\parindent}{2em}
       \affaddr{Beihang University}\\
       \setlength{\parindent}{3em}
       \email{wujj@buaa.edu.cn}
\alignauthor Guannan Liu\\
       \affaddr{Tsinghua Universtiy}\\
       \email{liugn.10@sem.tsinghua.edu.cn}
\and
\alignauthor Ke Xu\\
       \affaddr{Beihang University}\\
       \email{kexu@nlsde.buaa.edu.cn}
\alignauthor Guoqing Chen\\
       \affaddr{Tsinghua Universtiy}\\
       \email{chengq@sem.tsinghua.edu.cn}
}

\maketitle
\begin{abstract}
Recent years have witnessed the explosion of online social networks (OSNs). They
provide powerful IT-innovations for online social activities such as organizing contacts, publishing
contents, and sharing interests between friends who may never meet before. As more and more people become the active
users of online social networks, one may ponder questions such as: (1) Do OSNs indeed improve our sociability?
(2) To what extent can we expand our offline social spectrum in OSNs? (3) Can we identify some interesting user behaviors
in OSNs? Our work in this paper just aims to answer these interesting questions. To this end, we pay a revisit to the well-known Dunbar's number
in online social networks. Our main research contributions are as follows. First, to our best knowledge, our work is the first one that systematically
validates the existence of the online Dunbar's number in the range of [200,300]. To reach this, we combine using local-structure analysis and user-interaction
analysis for extensive real-world OSNs. Second, we divide OSNs users into two categories: rational and aggressive, and find that rational users intend to develop
close and reciprocated relationships, whereas aggressive users have no consistent behaviors. Third, we build a simple model to capture the constraints of time and
cognition that affect the evolution of online social networks. Finally, we show the potential use of our findings in viral marketing and privacy management in
online social networks.
\end{abstract}




\section{\label{sec:introduction}Introduction}

In this day and age, the online social sites, like
Facebook\footnote{http://www.facebook.com},
Livejournal\footnote{http://www.livejournal.com},
MySpace\footnote{http://www.myspace.com} and etc., provide people
with a new powerful means to communicate and interact with each
other. Through these sites, users can share blogs, photos and
current statuses. They can consolidate friendships in the real-world
by exchanging information online and establish new virtual
friendships with others in the same site. It is these sites that
lead to the formation of a new kind of social network, which is
called the online social network. Indeed, with the thorough
development of online social sites in the recent decade, the online
social network has become an essential part of our daily life and is
changing our social behaviors potentially. At the same time,
different from the traditional real-world social network, the
electric communication data of the online social network is
relatively easy to collect\cite{inferringsocialnetworks}. Besides,
compared with the real-world social network, its scale is huge. So
it is reasonable to conjecture that this new form of network would
give many inspirations to the previous recognition of the social
networks.

Through coupling the number of friendships and the size of the neocortex
in primates, Dunbar found humankind can only maintain as many as 150 friendships
effectively\cite{dunbarnumber}. And the number 150 is then called
the magic number in social networks.
In our experience of using online social networks, we can easily find
that some users have extremely large number of friends, much more than 150,
while others keep only an averaged level of friends. We believe it is the
online mechanisms that facilitate the formation of high-degree nodes,
since friends making is so convenient that only requires an invitation to
be added as a friend and an acceptance. Therefore we may cast doubts on
whether online social networks deviate from the constraint of Dunbar's number.
To verify the doubts, it is necessary to investigate the following questions:
\begin{itemize}
\setlength{\itemsep}{0pt}
\item
    Does there still exist a magic number in online social networks
    as Dunbar's number in the real-world social networks?
\item
    If it exists, what's its value and how it generates?
\item
    How it changes things?
\end{itemize}

In this paper, we aim to answer the above questions through the
analysis on several datasets coming from some online social sites.
By observing many local measures, we conclude that there exists a
new magic number pervasively, which is in the range of 200 and 300,
greater than 150 found previously in real-world social networks. We
also validate this by investigating the traces of interaction
between many users in online social sites. We find through our
observations that although the online social sites provide us many
easier ways to maintain online friendships effectively, there is
still an upper limit on the number of substantial and meaningful
friends. Furthermore, we believe that users can be distinguished by
the magic number and they exhibit different behaviors and attitudes,
respectively.

Given the fact that many current models cannot interpret
these phenomena, we present a new simple model to interpret
how the magic number in online social networks generates.
Finally, we think this number is insightful to guide the
viral marketing strategy and user privacy management. For
instance, hub nodes may not be effective choice in viral marketing,
and certain users call for a detailed privacy setting mechanism.

The rest of the paper is organized as follows. In
Section~\ref{sec:relatedworks}, some related works will be
introduced. In Section~\ref{sec:preliminary}, we will define some
local and global measures used in the following analysis. Our
observations and findings will be depicted in
Section~\ref{sec:anewmagicnumber}. In Section~\ref{sec:model}, we
present a new model to interpret the new upper limit existing
pervasively in online social networks. We also give a talk about the
business insights about the new magic number in
Section~\ref{sec:businessinsights}. Finally, we conclude this paper
in Section~\ref{sec:conclusion}.

\section{Related Work}
\label{sec:relatedworks} Our study is related to the work in three
areas: the phenomenon of Dunbar's number, measurement analysis of
online social networks and social network modeling.

\subsection{The phenomenon of Dunbar's number}
\label{sec:dunbarnumber}

By investigating the relationship between neocortex size and group
size in primates, Dunbar\cite{dunbarnumber} predicted the number of
group size in human beings was 150, which was notable as Dunbar's
number. According to him, human beings can only maintain a small
fraction of relationships within the circle of Dunbar's number, and
other relationships beyond that circle are not reciprocated or
personalized. Dunbar's number targets on real-world social networks
when first put forward. However, recent works in online social
networks have displayed similar interesting observations
\cite{ahn:analysis,golder:rhythms}. Roberts et al. pointed out that
time spent using social media, including online social sites, was
not associated with larger offline networks
\cite{offlinenetworksize}. Potential time and cognitive constraints
were also considered in their work. Other work of online social
networks related to Dunbar's number will be further discussed in
Section \ref{sec:measurementanlysis}.

\subsection{Measurement analysis of online social networks}
\label{sec:measurementanlysis}

Recently, researchers have done intensive study in online social
networks. They measured the property of online social networks from
different perspectives. Phenomena such as small world, power-law,
high clustering, assortativity have been observed in different
social sites, which are believed to be the common properties of
online social networks. Ahn et al. studied the largest online social
networks Cyworld in South Korea\cite{ahn:analysis}. They
experimented on the whole data of Cyworld and discovered some unique
characters of this site. They found an interesting phenomenon that
most user connections were not active and attributed it to Dunbar's
number. Mislove et al. used data from Flickr, YouTube, LiveJounal
and Orkut, conducting measurement analysis in a large
scale\cite{mislove:measure}. They incorporated various complex
network measurements such as degree distribution, clustering
coefficients, degree correlations, connected core etc. in the
research. Golder et al. analyzed Facebook users in North American
colleges or universities\cite{golder:rhythms}. Their results on
degree distribution showed that the number of people who have few
hundreds of friends remained stable, but it started to drop sharply
once the friends number exceeded 250, which also coincided with
Dunbar's number.

\subsection{Social networks modeling}
\label{sec:socnetmodel}

Although Small-World\cite{sm-model} and BA\cite{ba-model} network
models lay a foundation for the complex network, these two models
cannot explain all the phenomena of different types of real
networks. As for social networks, multiple models have been proposed
to fit their particular properties. Holme and Kim added a \lq\lq
triad formation step\rq\rq~beyond BA model(HK), \emph{i.e.}
establishing edges between neighbors of a node\cite{holme:growing}.
Davidsen et al. imported a similar process in their DEB
model\cite{davidsen:emergence}. This process in fact corresponds to
the real-world social network situation, as people are easily
introduced to meet friends of their friends and build up
connections. Both networks generated from HK and DEB have power-law
and high clustering properties. Jin et al. carefully studied the
principles of social network formation, and proposed a social
network model (JGN)\cite{jin:structure}. JGN also considered the
influence of mutual friends, cost of friend maintenances and the
maximum connections. JGN and HK are the two early models to generate
social networks, and they both grasp the core principle to generate
networks-\lq\lq transitivity\rq\rq. Thus many successive models
inherit the idea of \lq\lq transitivity\rq\rq~from them.
The models mentioned above mainly focus on the real-world social
networks, while online factors are not taken into account. Although
online social networks have gained popularity in recent research,
the work to model this kind of network is still quite insufficient.
Yuta et al. pointed out that the cost of online friend maintenance
was much lower, and they extended CCN\cite{vaz:growing} by adding a
process of random linkage to form a new model CCNR\cite{yuta:gap}.
Bonato et al. also adopted transitivity in their model
ILT\cite{bonato:dynamic}.

In summary, almost all the social network models, no matter online
or offline, adopt the rule of transitivity in various forms. And the
networks generated by these models commonly have the feature of high
clustering due to this fact.

\section{Preliminaries}
\label{sec:preliminary}

In this section, we depict definitions of some critical global and
local measures which we would use in the following sections.

An online social network can be intuitively modeled as a graph
$G(V,E)$, where $V$ is the set of users and $E$ is the set of ties.
For the reason that establishing a new tie usually needs mutual
permission in online social sites, $G$ is undirected. Generally, the
number of a node's friendships can be defined as its degree. The
averaged degree of the network can be defined as
\begin{equation}
\langle k \rangle=\frac{2|E|}{|V|}.
\end{equation}
$k_{max}$ is the maximum degree among all the nodes and $k_{min}$ is
the minimum degree. $p(k)$ is the degree distribution of the graph
and for online social networks, it is always power-law. Usually, the
complementary cumulative distribution function ($CCDF$) is used to
characterize this.

Clustering coefficient of a node is used to characterize how closely
its neighbors are connected. It can be defined as
\begin{equation}
C_i=\frac{2|E_i|}{k_i(k_i-1)},
\end{equation}
where $E_i$ is the set of ties between $i$'s neighbors and $k_i$ is
the degree of $i$. For the case of $k_i=1$, we set $C_i=0$ in this
paper. Then the averaged clustering coefficient of the nodes with
degree $k$ can be defined as
\begin{equation}
C(k)=\frac{\sum_{\{i \in V|k_i=k\}}{C_i}}{|\{i \in V|k_i=k\}|}.
\end{equation}
The averaged clustering coefficient of the network can be defined as
\begin{equation}
C=\frac{\sum_{\{i \in V\}}{C_i}}{|V|}.
\end{equation}
The averaged clustering coefficient of the social network is always
higher than the technical network.

The averaged degree of a node's neighbors, denoted as $k_{nn}$, is
always used to depict the assortativity of the network. If the
network is disassortative, the nodes with low degree is
preferentially connected to ones with high degrees, then $k_{nn}$
will decrease with the increment of the degree. Contrarily, the
nodes will be connected to those with similar degrees when the
network is assortative. The social network is usually thought to be
assortative. Here we define $i$'s $k_{nn}$ as
\begin{equation}
k_{nn}^{i}=\frac{\sum_{j \in \{i's~neighbors\}}k_j}{k_i}.
\end{equation}
Similarly, the averaged $k_{nn}$ of the nodes with degree $k$ can be
defined as
\begin{equation}
k_{nn}(k)=\frac{\sum_{\{i\in V|k_i=k\}{k_{nn}^{i}}}}{|\{i \in
V|k_i=k\}|}.
\end{equation}
It can be divided by $k_{max}$ to be normalized.

K-shell (k-core) index, denoted as $k_s$, is usually used to
characterize how far is a node away from the core of the network.
For instance, greater value of $k_s$ means the node is closer to the
core. It can be obtained through the following method
\cite{kshell-influential}. First, remove all the nodes with degree
$k=1$. After this stage of pruning, there may appear new nodes with
$k=1$. Then keep on pruning these nodes, as well, until all nodes
with degree $k=1$ are removed. $k_s$ of the removed nodes will be
set to 1. Next, we repeat the pruning process in a similar way for
the nodes with degree $k=2$ and subsequently for higher values of
$k$ until all nodes are removed. In \cite{kshell-influential}, it is
found that in many networks, including online social networks,
high-degree nodes may have low $k_s$, indicating that those nodes
were at the periphery of the network. The averaged k-shell index of
the nodes with degree $k$ can be defined as
\begin{equation}
k_s(k)=\frac{\sum_{\{i\in V|k_i=k\}}{k_s^i}}{|\{i\in V|k_i=k\}|},
\end{equation}
where $k_s^i$ is the k-shell index of $i$.

The strength of a tie between two nodes in a social network is
usually defined as the overlap of their friends \cite{finding-job,
mobile-network}. It means the more common friends they share, the
more familiar they would be. In online social networks, sharing more
common friends usually means they are geographically close to each
other, or share the same profiles, or interact more frequently
online. Online friends with a big value of tie strength may have a
higher probability to be friends in offline social networks. We
define the strength of tie between $i$ and $j$ as
\begin{equation}
w_{ij}=\frac{c_{ij}}{k_i-1+k_j-1-c_{ij}},
\end{equation}
where $c_{ij}$ is the number of common friends between node $i$ and
$j$, $k_i$ and $k_j$ is the degree of $i$ and $j$, respectively.
Based on the definition of tie strength, we can also define the
strength of a node as the averaged strength of all the ties
connected to it. It is
\begin{equation}
w_i=\frac{\sum_{j\in \{i'~neighbors\}}w_{ij}}{k_i}.
\end{equation}
Then we can define the averaged strength of the nodes with degree
$k$ as
\begin{equation}
w(k)=\frac{\sum_{\{i\in V|k_i=k\}}{w_i}}{|\{i\in V|k_i=k\}|}.
\end{equation}

\section{A New Magic Number}
\label{sec:anewmagicnumber}

In this section, we start from some observations on the local
measures of a sampled online social network from Facebook. By
coupling the variation of local measures with the increment of the
degree, we discover an interesting phenomenon.
Then we find this phenomenon pervasively exists in other online
social networks. We summarize these observations to formulate our
conjectures. In the end of this section, we also validate them by
investigating the real trace of online interactions between users.

\subsection{The turning point on local measures}
\label{sec:inflectionpoint}

The sample dataset we use comes from \cite{viswanath-2009-activity}
and it is publicly available. This dataset is a snapshot of Facebook
network in the city of New Orleans, so we denote it as
\texttt{NewOrleans}. It contains 63292 nodes and 816886 ties. Its
$k_{max}=1098$, $k_{min}=1$, $\langle k \rangle=25.8$ and $C=0.22$.

\begin{figure}[ht]
\centering \epsfig{file=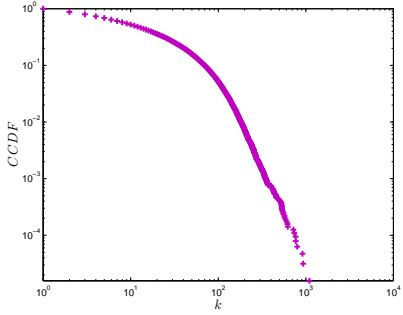,width=6cm}
\caption{Degree distribution of \texttt{NewOrleans}}
\label{fig:degdist}
\end{figure}

We firstly analyze the measurement of degree distribution for the
\texttt{NewOrleans} network. As shown in Figure~\ref{fig:degdist},
we surprisingly find a gentle slope in the interval between [0,200]
in the degree distribution. Unlike a straight line in typical
power-law, an turning point obviously appears and power-law only
exists in the tail.

We calculate the fraction of users in the gentle slope interval and find
more than 94\% users are in it. Why most users are so \lq\lq
crowded\rq\rq~in this narrow interval while the number of users
begin to drop dramatically beyond the interval?

In fact, similar degree distribution has already been observed by
Golder et al. in \cite{golder:rhythms}. The turning point in their
dataset approaches 250, and they argue that it is because friendship
in \texttt{Facebook} cannot completely represent conventional
friendship. Nevertheless, they did not further explore the problem
to give a more detailed explanation.

To sum up, node degree $k\in[200,300]$ denote a threshold value, and
distributions are different in the two sides of it. By revealing the
threshold, we may want to know whether there are any hidden facts
lay behind it. To recall Dunbar's number mentioned in Section
~\ref{sec:dunbarnumber}, we can see the threshold is not far from
Dunbar's number 150. So does Dunbar's number play a vital role in
this phenomenon? Does it exist or shift to a new magic number in
online social networks? To provide more concrete evidences to
explain the phenomenon, we observe how $C(k)$, $k_{nn}(k)$ and
$w(k)$ related with $k$ in this network to see whether turning point
also appears in these measurements, and we conclude our observations
and remarks as follows.

\begin{figure*}[ht]
\centering
 \subfigure[\texttt{$C(k)$}]{\epsfig{file=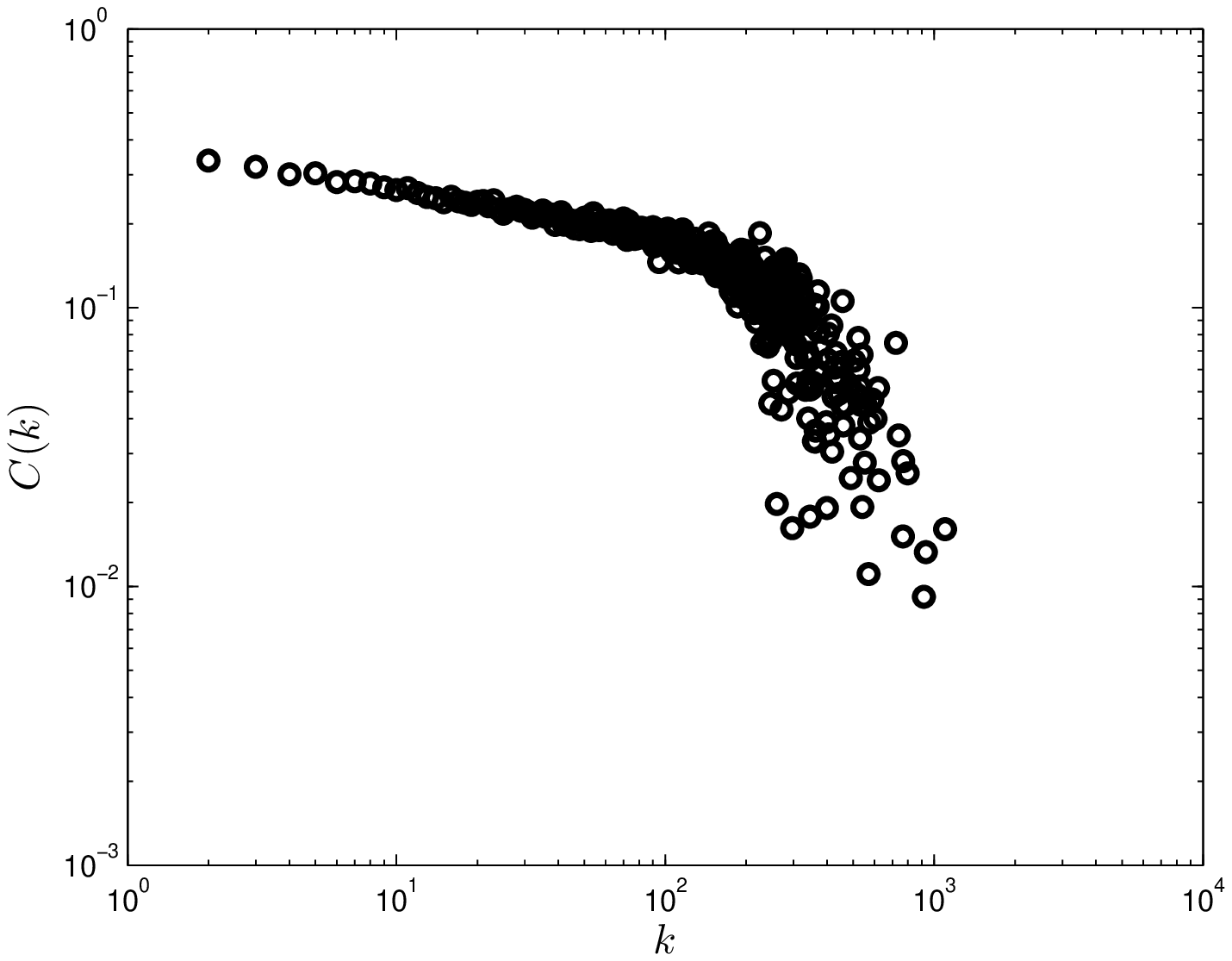,width=5.2cm}\label{fig:neworleans-dmapac}}
 \subfigure[\texttt{$k_{nn}(k)$}]{\epsfig{file=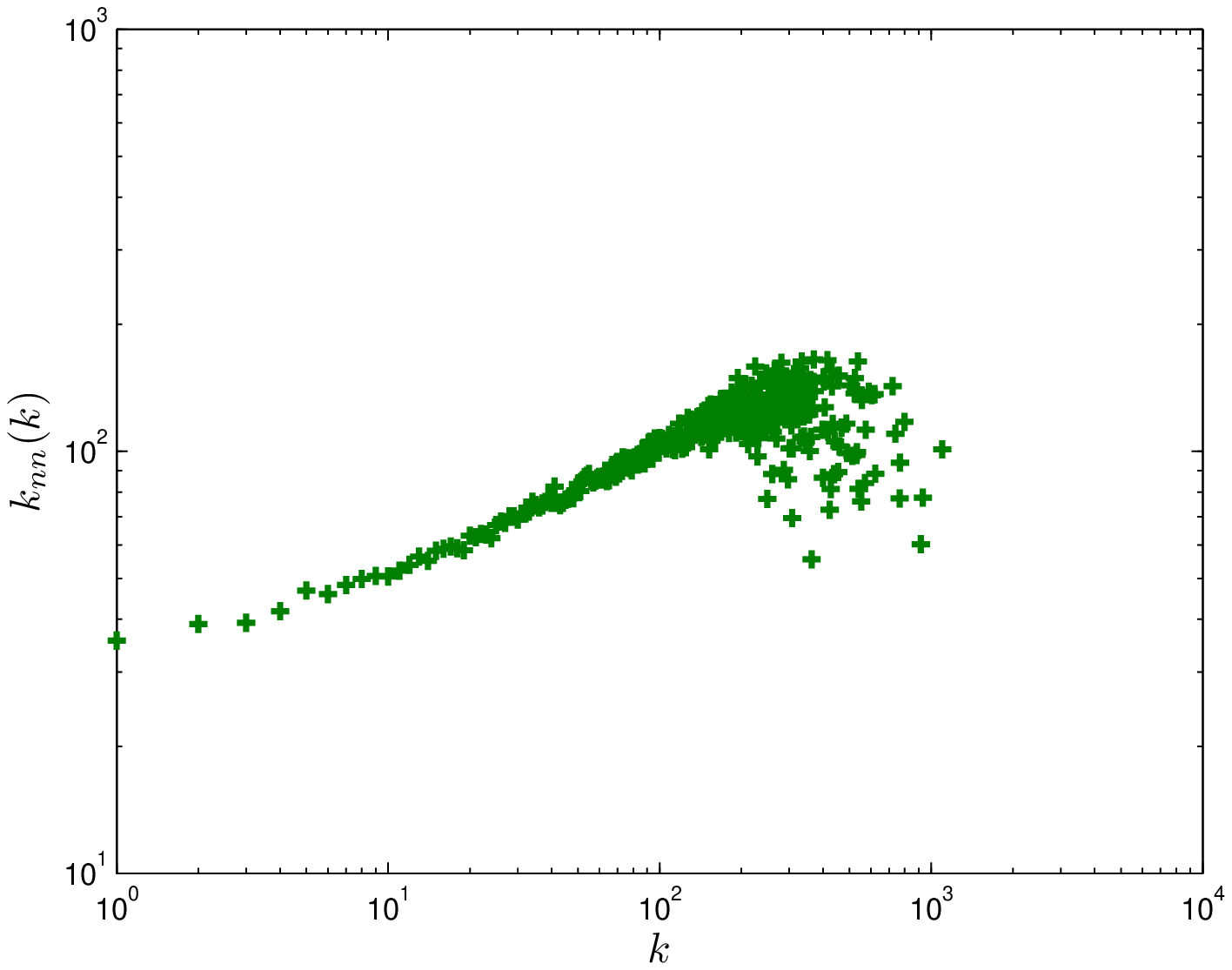,width=5.2cm}\label{fig:neworleans-dmapknn}}
 \subfigure[\texttt{$w(k)$}]{\epsfig{file=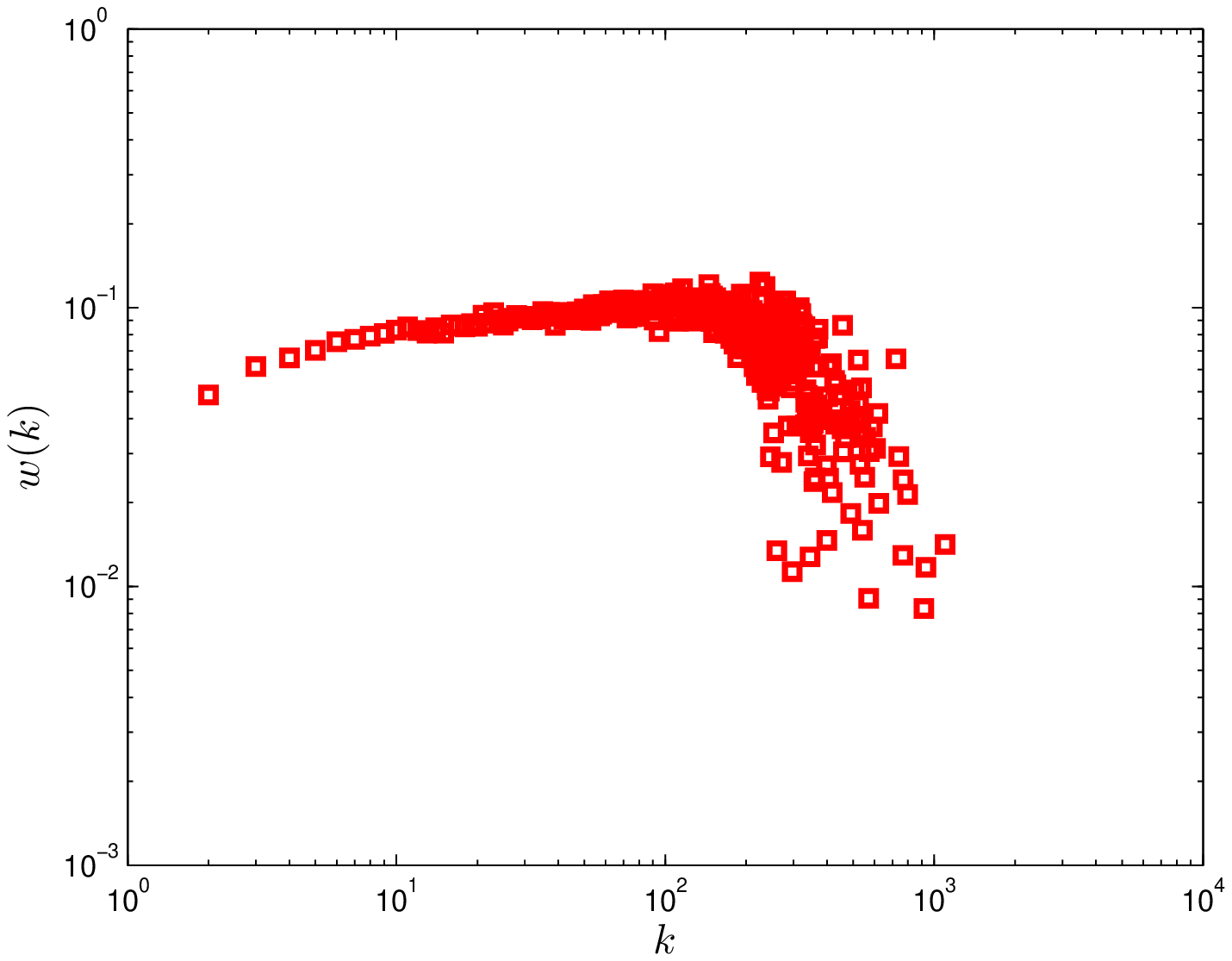,width=5.2cm}\label{fig:neworleans-dmapns}}
\caption{The variations of local measures for \texttt{NewOrleans}.}
\label{fig:neworleans}
\end{figure*}

{\bf Observation 1.} We plot the distribution of clustering
coefficient, as shown in Figure~\ref{fig:neworleans-dmapac}. $C(k)$
steadily decreases with the increment of $k$ at first, however, when
$k$ exceeds a certain threshold in the region of [200,300], the
speed of decrement apparently raises. Therefore the turning point
also exists in $C(k)$ with almost the same value in the degree
distribution. Here we simply denote it as $k_T$ and $k_T \in
[200,300]$.

{\bf Remark 1.} Clustering coefficient reflects the connections
among neighbors of a node. High clustering coefficient indicates
tightly connected neighborhood. In view of {\bf Observation 1}, we
can conclude that users with lower degrees($k<k_T$) have a well
connected neighborhood, \textit{i.e.,} quite a large fraction of
these users' friends are acquainted with each other. It's not hard
to explain this in the real social networks, as a person always get
acquainted with some strangers through one of his friends. The
behavior of meeting friends' friends is even strengthened in online
social networks since it enables users to meet others with no
restriction in time and space. For instance, Facebook users will
receive notifications in their \emph{News Feed} when their friends
establish friendship with another user, and they can click \lq\lq
add as friend\rq\rq~to become friends as well. Moreover, almost all
the social sites provide the feature of \emph{Common Friends}, which
lists the other users having common friends with you. It is this
form of friends making mechanisms that leads to a denser network for
lower-degree users. On the contrary, things cannot hold true for the
users with high degrees. The averaged clustering coefficient drops
to a very low point, meaning that although these users have hundreds
of friends, they do not know them well indeed. In consequence, we
believe that most friends of these users seldom make new friends by
way of them, making them have loose neighborhoods. It seems that
users are divided into two types by the turning point $k_T$, with
one type of users positioned in an acquaintance network and
maintaining some meaningful connections, and the other type of users
keeping some formalized relationships.

{\bf Observation 2.} Figure~\ref{fig:neworleans-dmapknn} displays
the distribution of degree correlations, the trend of which can
represent the assortativity of the network. It's interesting to find
there is a positive trend within $k_T$, and then $k_{nn}(k)$
scatters with a slight negative trend beyond that. The same trend of
$k_{nn}(k)$ has also been observed in Mixi, an online social site in
Japan \cite{yuta:gap}.

{\bf Remark 2.} The positive trend of $k_{nn}(k)$ at the first stage
is consistent with the assortativity property of the offline social
networks \cite{newman-cluster}. In this stage, nodes tend to connect
to those with similar degrees. However, the negative scattering in
the range $[200,300]$ shows that the network transforms to a
disassortative network. That is to say, for the users with degrees
higher than $k_T$, they are not preferentially connected to the
nodes with similar degrees. Results in {\bf Observation 2} suggest
that users on the two sides of the turning point behave differently
in establishing friendships. On the left side, users are prone to
connect other users with similar number of friends, while users on
the right side may randomly add large amount of friends without much
consideration. The results from degree correlations provide another
evidence to prove the distinction of users in online social
networks.

{\bf Observation 3.} It is easy to find that there exists an turning
point in Figure~\ref{fig:neworleans-dmapns}, still within the value
$k_T$. Before reaching the turning point, $w(k)$ steadily remains a
high value with slight increment as $k$ grows, but it begins to
decrease when $k$ is higher than $k_T$.

{\bf Remark 3.}  The previous work \cite{mobile-network} has shown
that in social networks, like mobile communication network, the tie
strength $w_{ij}$ between node $i$ and $j$ is strongly related with
the frequency of interaction between users. It is also found that in
online social networks, nodes with more mutual friends tend to trust
each other and share more similar interests \cite{mislove:measure}.
So the averaged tie strength $w_i$ of user $i$ can imply the overall
quality of relationships with friends to some extent. View from {\bf
Observation 3}, users with degrees lower than $k_T$ keep a high tie
strength value, suggesting that these users maintain friendships
with high quality. In contrast, for users with degrees higher than
$k_T$, their strength is weak and become even weaker with the
increment of $k$. Then we infer that among their friendships, some
are fragile and not trusted. In the online social sites, this
situation possibly corresponds to the following scenarios:
\begin{itemize}
\setlength{\itemsep}{-1pt}
\item
    High-degree users are really popular to attract a lot of
    low-degree nodes to add them as friends. For example, they can
    be a movie star, a notable scientist or even a famous enterprise, etc.
    However, they do not communicate with those \lq\lq \emph{new}\rq\rq~friends
    frequently. Therefore the ties between them become weak and some
    may even vanish eventually.
\item
    It's human nature to pursue for prestige in the society, while
    the forms may be different. Some people are eager to gain popularity
    by randomly sending thousands of invitations to be added as friends
    in online social networks. They can probably acquire thousands of
    online friends as time goes by, however, most of the relationships
    are of no meaning and the number of shared friends is certainly
    quite small.
\end{itemize}

We surprisingly find that the same turning point appears in all the
measurements, coinciding with what we've found in the degree
distribution. Based on the above observations and remarks, we can
conjecture that in the dataset we used, there does exist a threshold
$k_T\in[200,300]$. As a matter of fact, this magic threshold
distinguish users by their variations of the local topological
properties in online social networks. Furthermore, we have found
that users' online behaviors can be characterized by these
properties.

As Dunbar's number implies, an individual could not maintain more
than 150 friendships in the real world effectively because of the
time and cognitive constraints. We unveil in the previous
observations that users beyond the turning point behave quite
differently. They keep loose neighborhood with only a few friends
knowing each other, and they randomly connect to users of different
degrees or demographics with no preference; more importantly, the
weak averaged tie strength of these users indicates poor
relationships with their friends. It seems that the turning point
$k_T$ plays a similar role in online social networks as Dunbar's
number in the real society due to these facts. Although it is at low
cost to make new friends and maintain friendships in online social
networks, the number of friendships one can handle is still limited.
In view of this, we infer that the turning point $k_T$ is just the
upper limit of well maintained online friendships, and users can be
divided into two categories based on this point.

Online social networks has gained so much popularity in the
worldwide. However, people's attitude towards it gradually changes
with more intensive use. In this day and age, logging in your
Facebook account is not purely for entertainment but transforms into
a habit and everyday life, just like checking out your emails or
browsing a web page. Many users become rational in using social
sites, as they carefully maintain a well connected neighborhood,
most of which are \lq\lq cloned\rq\rq~from their offline social
networks. So the motives of using online social networks for the so
called \lq\lq rational users\rq\rq~become as simple as maintaining
friendships. They stick to this new form of friendship maintenance
because it shortens the distance between friends, as they can be
informed of what is happening to their friends through the news feed
with no geographic constraints. In fact, \lq\lq rational
users\rq\rq~corresponds to the users within $k_T$. Nonetheless,
users beyond the threshold denote another type of users. We define
them as \lq\lq aggressive users\rq\rq~as they aggressively
accumulate large amount of friends while most relationships are
inactive and lack of interactions.

However, we draw our conclusions only from one dataset of Facebook
so far. Does the magic threshold exist in other online social
network samples? Or if it exists, do users behave distinctively on
the two sides of the threshold? In the next section, we employ more
datasets, either larger or smaller, to further discuss these issues.

\subsection{Pervasiveness}
\label{sec:pervasiveness}

In this section, in order to prove the ubiquity of the phenomena we
found above, we import another five datasets of online social
networks. The first four datasets are provided by
\cite{facebook-community} and are all publicly available. The four
datasets are the complete Facebook networks whose ties are within
four American universities. The four universities are Georgetown
University(\texttt{Georgetown}), Princeton
University(\texttt{Princeton}), University of
Oklahoma(\texttt{Oklahoma}) and University of North Carolina at
Chapel Hill(\texttt{UNC}), respectively. The fifth data set was
collected from Livejournal, denoted as \texttt{Livejournal}. It is
also public to the research community\cite{mislove:measure}. The
detailed descriptions of these datasets are listed in Table
~\ref{tab:datasets}.
\begin{table}
\centering \caption{Datasets} \label{tab:datasets}
\begin{tabular}{|c|c|c|c|c|c|}
\hline
Dataset & $|V|$ & $|E|$ & $k_{max}$ & $\langle k \rangle$ & $C$ \\
\hline
\texttt{Georgetown} & 9388 & 425619 & 1235 &  90.67 & 0.22\\
\hline
\texttt{Oklahoma} & 17420 & 892524 & 2568 & 102.47 & 0.23\\
\hline
\texttt{Princeton} & 6575 & 293307 & 628 & 89.22 & 0.24\\
\hline
\texttt{UNC} & 18158 & 766796 & 3795 & 84.46 & 0.20\\
\hline
\texttt{Livejournal} & 5203764 & 48709773 & 15017 & 18.72 & 0.27\\
\hline
\end{tabular}
\end{table}

\begin{figure*}[ht]
\centering
 \subfigure[\texttt{Degree Distribution}]{\epsfig{file=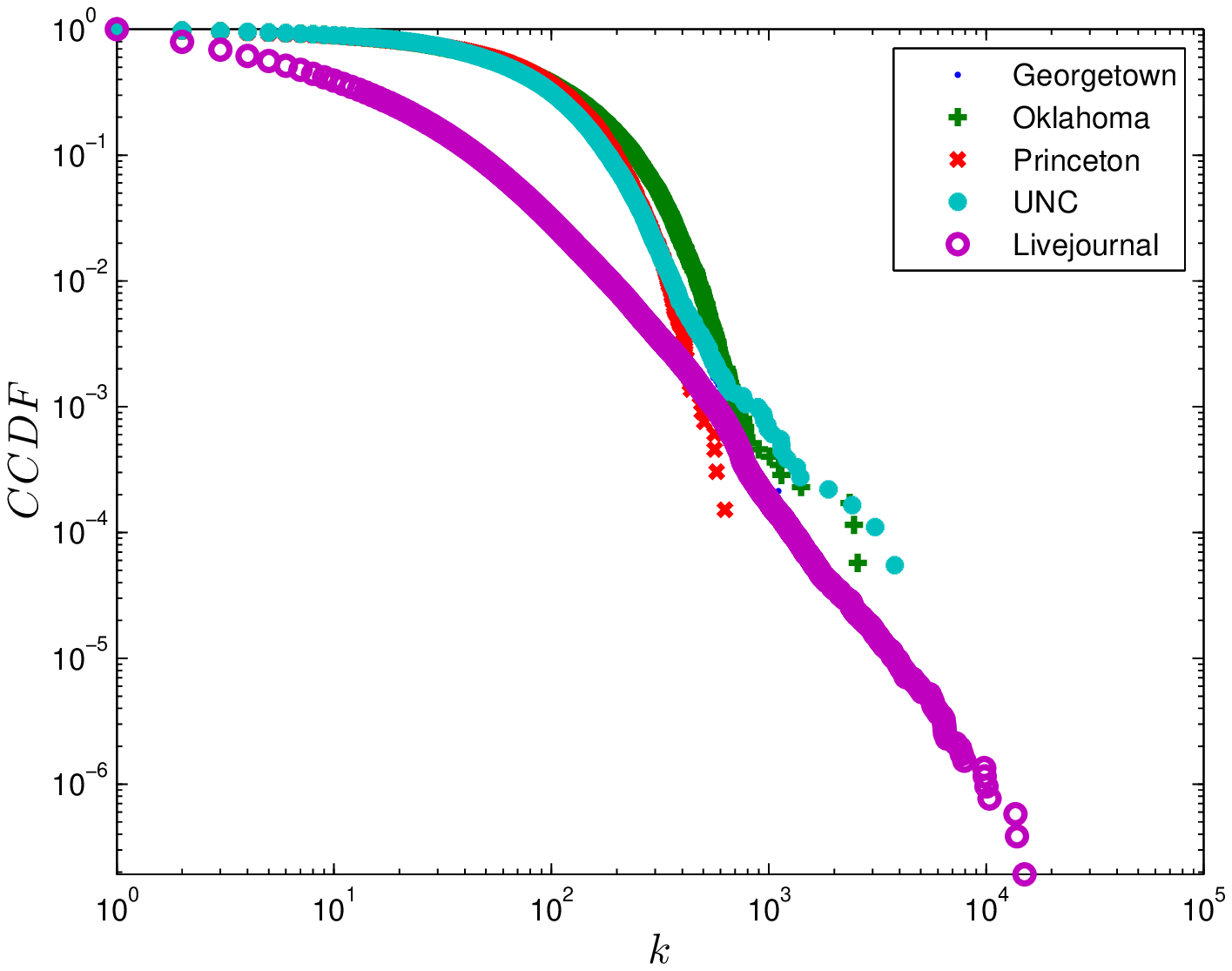,width=5.2cm}\label{fig:georgetown}}
 \subfigure[\texttt{Georgetown}]{\epsfig{file=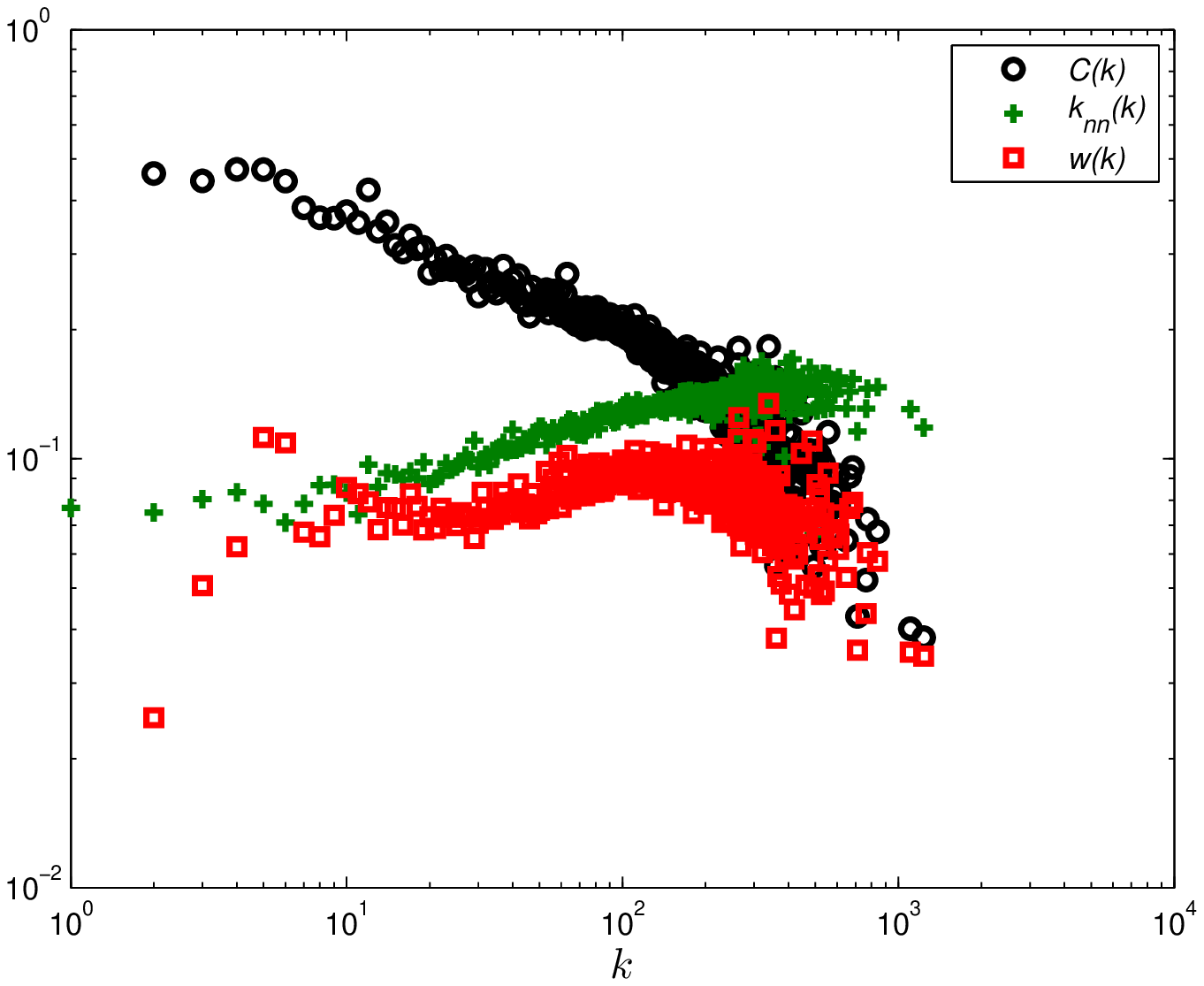,width=5.2cm}\label{fig:georgetown}}
 \subfigure[\texttt{Oklahoma}]{\epsfig{file=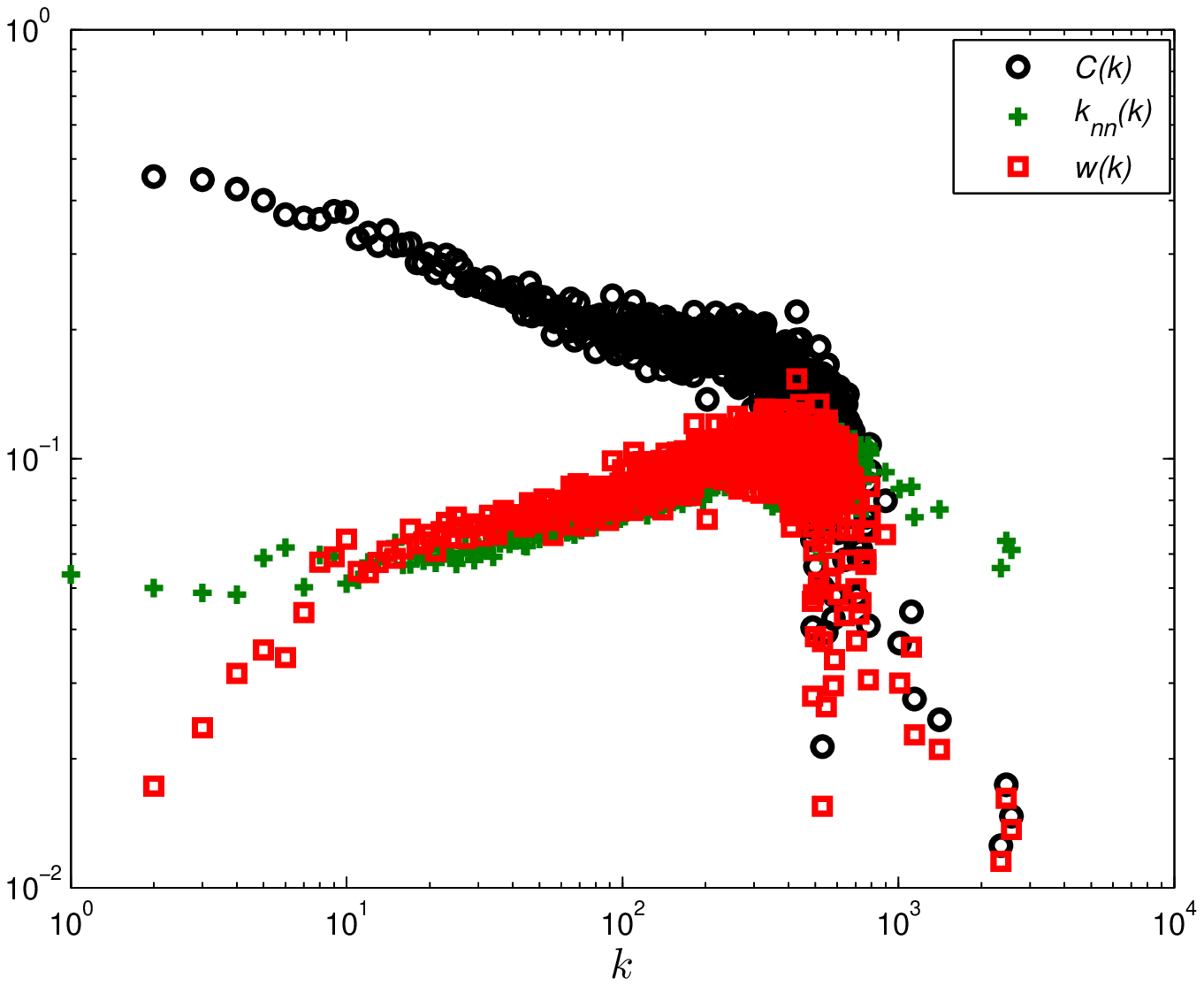,width=5.2cm}\label{fig:oklahoma}}
 \subfigure[\texttt{Princeton}]{\epsfig{file=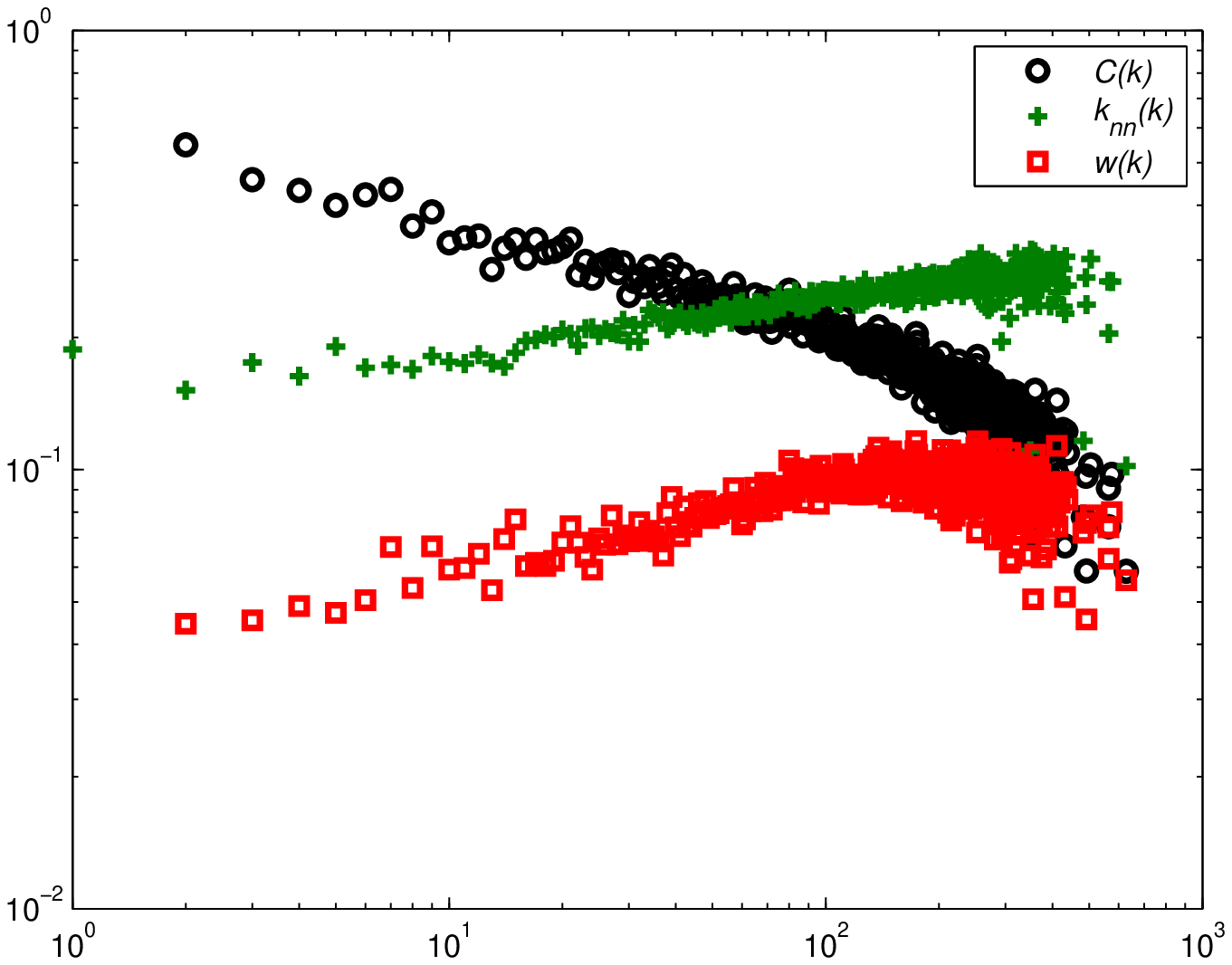,width=5.2cm}\label{fig:princeton}}
 \subfigure[\texttt{UNC}]{\epsfig{file=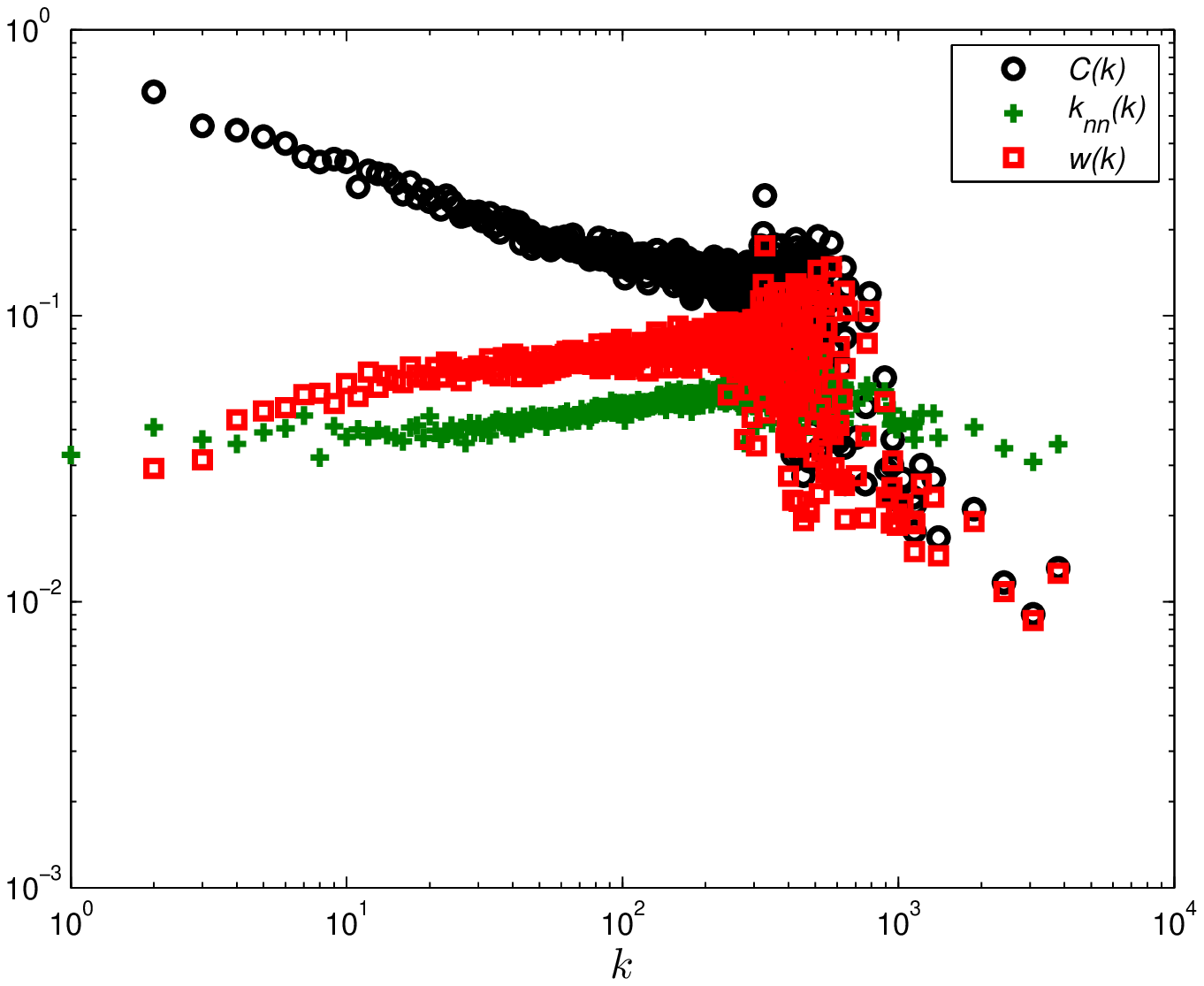,width=5.2cm}\label{fig:unc}}
 \subfigure[\texttt{Livejournal}]{\epsfig{file=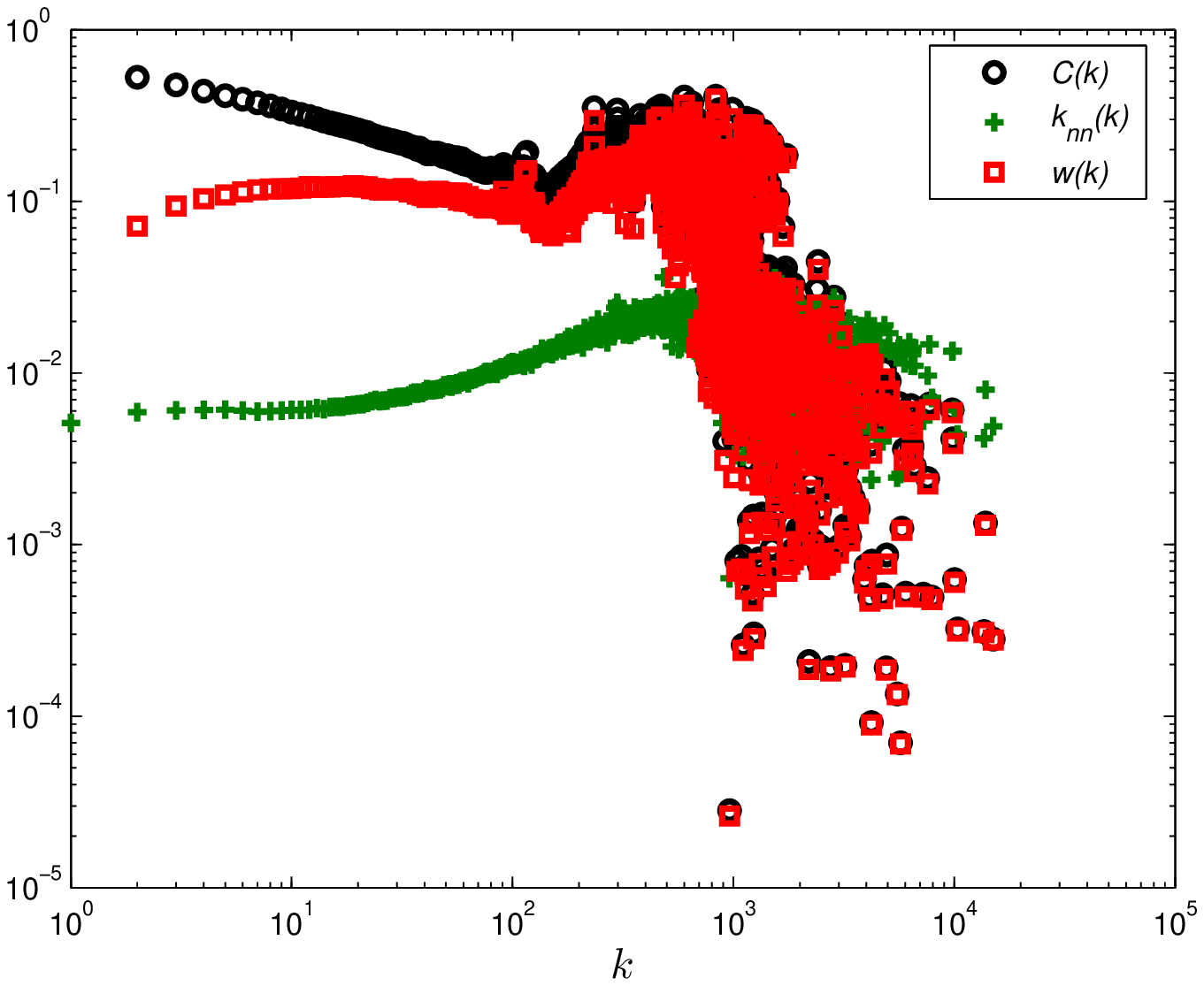,width=5.2cm}\label{fig:livejournal}}
\caption{Results from other datasets.} \label{fig:otherdatasets}
\end{figure*}

Next, we perform the same measurements on these five datasets as on
\texttt{NewOrleans}. Just as shown in
Figure~\ref{fig:otherdatasets}, for all the measures, including
$CCDF$, $C(k)$, $k_{nn}(k)$ and $w(k)$, there still exists a
threshold $k_T\in[200,300]$, which is independent to $|V|$.
Especially, in the dataset of \texttt{Livejournal}, the size of the
network is as many as five millions, however, the $k_T$ is still in
the range between 200 and 300.

In addition, as users' demographic data is provided in the datasets
of \texttt{Georgetown}, \texttt{Oklahoma}, \texttt{Princeton} and
\texttt{UNC}, we can conduct experiment to examine the
homophily\cite{homophily} property in the network.  In these
datasets, we investigate the homophily in the following contexts,
including student or faculty flag, gender, major, second major,
dorm, year of enrolling and high school. The attribute vector for a
node can be defined as
$Hi(a_1^i,a_2^i,a_3^i,a_4^i,a_5^i,a_6^i,a_6^i,a_7^i)$, where
$a_l^i(l=1,2,...,7)$ corresponds to the above properties
sequentially. we define binary distance for each attribute , which
means, $\parallel a_l^i-a_l^j\parallel=1$ when $a_l^i\neq a_l^j$,
otherwise, $\parallel a_l^i-a_l^j\parallel=0$. Then the homophily
distance between node $i$ and $j$ can be defined simply as
\begin{equation}
d_{ij}=\parallel H_i-H_j\parallel
_2=\sqrt{\sum_{l=1}^{7}||a_l^i-a_l^j||^2}.
\end{equation}
The averaged homophily distance for node $i$ can be defined as
\begin{equation}
d_i=\frac{\sum_{j\in \{i's~neighbors\}}{d_{ij}}}{k_i},
\end{equation}
where $k_i$ is degree of $i$. Then the averaged homophily distance
for the nodes with degree $k$ can be defined as
\begin{equation}
H(k)=\frac{\sum_{\{i\in V|k_i=k\}}d_i}{|\{i\in V|k_i=k\}|}.
\end{equation}

\begin{figure}[ht]
\centering \epsfig{file=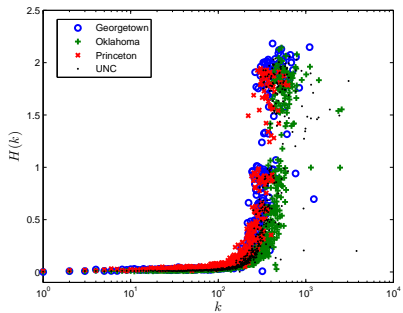, width=6cm}
\caption{Homophily property.} \label{fig:homophily}
\end{figure}

As shown in Figure~\ref{fig:homophily}, there is an explosive
increment of $H(k)$ beyond the same degree value $k_T$. As $H(k)$
measures the similarity between users' attributes, it convinces our
inference that \lq\lq rational users\rq\rq~have an tendency to
establish friendships with their familiar friends in the real world,
usually sharing some common demographics such as dorm, major etc.
While \lq\lq aggressive users\rq\rq~are somewhat aimless to add as
many friends as possible, so the demographics vary a lot.

{\bf Remark} We confirm that our findings from the dataset of
\texttt{NewOrleans} are pervasively existing in online social
networks through the above observations. We even strengthen our
conjectures that many connections of \lq\lq aggressive
users\rq\rq~are established with no substantial meaning by observing
the homophily property. That is to say, threshold phenomenon is not
a special feature of the sampled dataset, but a general character of
online social networks.

All the previous analysis, based on the six datasets, mainly relies
on the topological feature of the networks. However, pure structural
information cannot be so convinced to represent the interaction
between users. So it is necessary to validate our former conjectures
through some real-world traces of interactions, which will be
introduced in the next subsection.

\subsection{Validation}
\label{sec:validation}

Generally, the interaction data flowing in the online social sites
is hard to collect. Because there are always some configurations of
privacy protection in these sites. To our best knowledge, we collect
two datasets from \cite{viswanath-2009-activity} and
\cite{applicationdata}, respectively. Both of the datasets are
publicly available and anonymized for research purpose. Besides, the
two datasets both come from Facebook.

The first dataset was collected from the Facebook network in the
city of New Orleans, which was related to the dataset of
\texttt{NewOrleans} mentioned in the previous sections. They
collected the publicly accessible profile pages and abstracted the
list of the \lq\lq \emph{Wall}\rq\rq~post. So we denote this dataset
as \texttt{NewOrleans-Wall}. \lq\lq \emph{Wall}\rq\rq~is a popular
feature of Facebook, through which a user can leave messages on his
friends' profile pages and the friends can also reply him by leaving
messages, too. It is a classical and easy interacting way in
Facebook. The dataset covers as many as 46952 users.

The second dataset was collected through some Facebook applications
developed by the authors of \cite{applicationdata}, denoted as
\texttt{Facebook-Applications}. They developed three popular
Facebook applications named as GotLove?, HUG and Fighters' Club,
respectively. The authors collected about 3-week traces starting
from March 20, 2008. Here we only use the data from
\texttt{GotLove?} and \texttt{HUG}. In \texttt{GotLove?}, one node
can send \lq love\rq~to its friends. And in \texttt{HUG}, a user A
can send a virtual \lq hug\rq~to a friend B. The dataset from
\texttt{GotLove?} contains 642088 active users and the one from
\texttt{HUG} contains 198379 active users.

Based on these traces of interaction in Facebook, we try to relate
the degree of nodes with their activity strength and then to
validate our previous findings. In the dataset of
\texttt{NewOrleans-Wall}, we define the node $i$'s activity strength
as the length of its list of the wall post, which can be denoted as
$L_i$. The longer the list of the wall post is, the more
interactions between $i$ and its neighbors happen. Then the averaged
activity strength of the nodes with degree $k$ can be defined as
\begin{equation}
L(k)=\frac{\sum_{\{i\in V|k_i=k\}}L_i}{|\{i\in V|k_i=k\}|}.
\end{equation}

For the dataset of \texttt{Facebook-Applications},
we simply extract the number of \lq love\rq~or \lq hug\rq~ the node
$i$ has sent ($s_i$) and received ($\gamma_i$). Then the
reciprocation of a node $i$, denoted as $r_i$, can be defined as the
ratio of $\gamma_i$ and $s_i$, \textit{i.e.}, $r_i=\gamma_i/s_i$.
Then the averaged reciprocation of the nodes with degree $k$ can be
defined as

\begin{equation}
r(k)=\frac{\sum_{\{i\in V|k_i=k\}}r_i}{|\{i\in V|k_i=k\}|}.
\end{equation}

\begin{figure}[ht]
\centering \epsfig{file=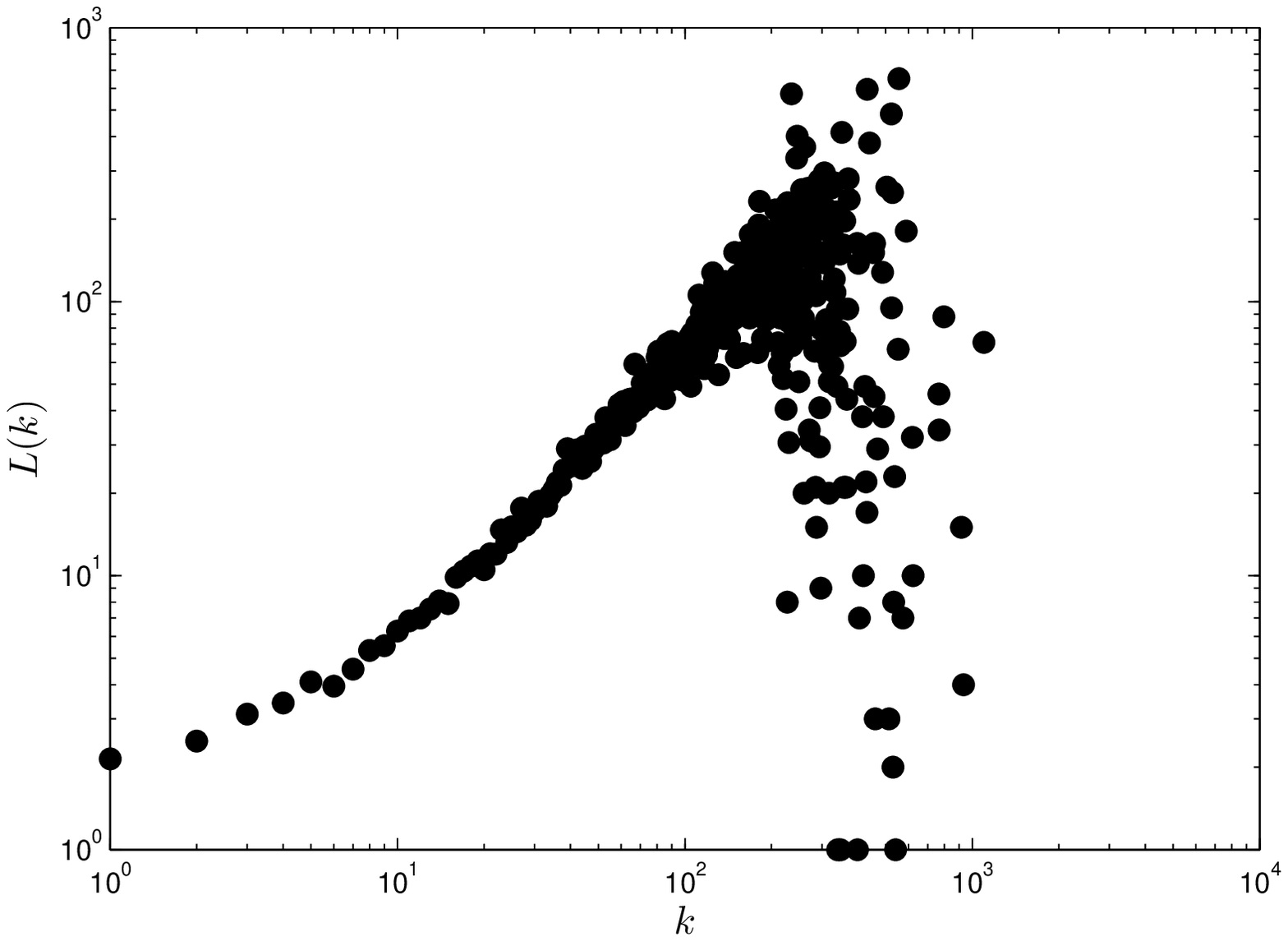, width=6cm}
\caption{\texttt{NewOrleans-Wall}} \label{fig:wall}
\end{figure}

{\bf Observation 1.} Just as shown in Figure~\ref{fig:wall}, as $k$
increases, $L(k)$ increases quickly. However, when $k$ reaches out
of the range [200,300], $L(k)$ stops increasing and begin to
fluctuate. This is consistent with our former finding from the pure
topological data. The fluctuation of $L(k)$ implies that some of the
users with degrees higher than the threshold $k_T$ have shorter list
of the wall post and their interaction remains in a rather low
level.

\begin{figure}[ht]
\centering \epsfig{file=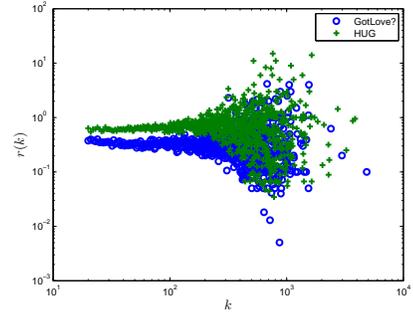,width=6cm}
\caption{Reciprocation varies with $k$.} \label{fig:reciprocation}
\end{figure}

{\bf Observation 2.} It is obviously shown in
Figure~\ref{fig:reciprocation} that the reciprocation of the nodes
approaches 1 when $k<k_T$. It means almost all the lower-degree
users' sent \lq love\rq~and \lq hug\rq~ are reciprocated. However
$r(k)$ diverges when $k>k_T$, indicating that their interaction with
friends is not symmetric. As has been illustrated in
Section~\ref{sec:inflectionpoint}, users accumulate many friends
either because of popularity or eagerness for prestige, thus we can
infer that some are far below 1 since these users' behaviors are
ignored by their friends, while some are above 1 because they are
popular enough to receive \lq love\rq~and \lq hug\rq~ from many
fans.

{\bf Remark} The above experiments further validate our conjecture
that there exists a degree threshold in online social networks. If a
user's degree is higher than the threshold, the user then cannot
maintain all its online friendships well and part of the friendships
can be easily ignored by the friends on the another end.

{\bf Summary} Until now, by validating from the real-world traces of
user interactions online, we can reasonably conclude that there
still exists an upper limit on the number of the friendships in
online social networks as Dunbar's number in offline social
networks. If users have more friends than the limit, it is
impossible for them to treat each tie equally. Because of this,
extraordinary dynamics will be bred when the degree goes up to the
limit. As a result, we see the phenomenon that high-degree users
keep overall relationships of low quality. In a further step, we
believe that users with mediate degrees are \lq\lq rational
users\rq\rq~with the motivation to maintain old friendships while
users who have friends more than the limit are likely to be
\lq\lq~aggressive users\rq\rq~seeking for new friends always.

However, little attention has been paid to these phenomena in many
current models, and they could not interpret the generation rule
completely. So we aim to understand how it generates by presenting a
new model in the next section.

\section{A New Model}
\label{sec:model}

In this section, we present a new model to interpret the generation
of the upper limit found in the previous sections. We start from
summarizing  users' online behaviors from previous observations and
conclusions. Next we introduce an inspiring model, and point out the
imperfections to apply the model to the situation of online social
networks. Then we incorporate the characters of users' online
behaviors to propose the new model. At last we examine the
properties of our simulated network to compare with the real
networks.

As has been discussed in Section ~\ref{sec:anewmagicnumber}, users'
online friends adding follows these rules:
\begin{enumerate}
\setlength{\itemsep}{-1pt}
\item
    When users first register in the sites, they tend to search for their offline acquaintances,
    and the network system would also recommend some friends based on the user's profiles.
    These friends become their initial online contacts, and they provide the basis for further friend making.
\item
    Even more conveniently than that in the real world, users can set up connections with their
    friends' friends simply by viewing the friends list and choosing who they already know to add as friends.
    Besides, some sites also recommend friends' friends to help users find more friends online.
\item
    As is illustrated in previous sections, most users are \lq\lq rational users\rq\rq~to be trapped in a magic number circle.
    After the number of friends is accumulated to a certain upper limit, they would stop adding more friends; or what's worse,
    they may even reject others' invitation to be added as friends.
    While only a few of sociable users jump out of the circle to become
    \lq\lq aggressive users\rq\rq, they'd like to add as many friends as possible
    actively. In fact, this process results in random linkage since the
    \lq\lq aggressive users\rq\rq~do not have explicit target users to link to.
\item
    Though online friendship maintenance costs almost no money or time, \lq\lq unfriend\rq\rq~situation still exists.
    For example, Section~\ref{sec:inflectionpoint} implies that the ties between \lq\lq aggressive users\rq\rq~and their
    friends are fragile and can vanish in some way. Moreover, social sites like Facebook will \lq\lq pull\rq\rq~all friends'
    updates to the users' news feed; however, some users may be annoyed by continually receiving one specific friend's message,
    thus \lq\lq unfriend\rq\rq~happens under this circumstance and the links are removed.
\end{enumerate}

In order to model the growth of social networks, Jin et al.
\cite{jin:structure} proposed two models based on three general
principles. First, the individuals tend to meet with those ones who
have one or more commons friends with them. Second, acquaintances
between the individuals who rarely meet decay over time, which means
some ties may vanish. Third, there is a maximum degree limit for  an
individual. However, features of the online social network are
different from their assumptions. For instance, the online social
networks usually starts to evolve from a real-world social network.
The site will urge the user to invite their real-world friends to
the site or provide easy way to search them in the site. The another
difference is the limit of the degree. In their models, a node can
not have a higher degree than the maximum degree. But in online
social networks, the maximum value set by the site may be high, like
5000 in Facebook or even more. Because of this, their model can only
control the maximum degree of the nodes, but can not interpret the
threshold degree as we find. In addition, this is also worth to be
noticed that in their model, the constraints of time and cognition
are only associated with the control of maximum degree. Given these
imperfections, we try to model the online social network based on
the following assumptions:
\begin{itemize}
\setlength{\itemsep}{-1pt}
\item
    The network start to evolve from a existing social network.
    Here, we simply start it from a BA network. The network evolves
    only by adding or removing ties.
\item
    New ties between nodes with common friends are preferred.
    However, for the nodes with degrees higher than a threshold,
    the probability of tie be established between its neighbors is
    lower.
\item
    Some nodes may search and add a random node as a friend.
\item
    Some ties may vanish, especially for the nodes with high
    degrees.
\end{itemize}
Guided by these principles, we present a simple model, called
$BA-shift$ as follows:
\begin{itemize}
\setlength{\itemsep}{-1pt}
\item
    Step 1: load a BA network, denoted as $BA(V,E_0)$, where $V$ is the
    set of nodes and $E_0$ is the set of original ties.
\item
    Step 2: In each time unit, perform the following actions:
    \begin{itemize}
    \setlength{\itemsep}{-1pt}
    \item
        Action 1: Select a node $i$ with the probability
        \begin{equation}
        p(i)=\frac{k_i(k_i-1)f(k_i)}{\sum_{j\in
        V}{k_j(k_j-1)f(k_j)}},
        \end{equation}
        where $k_i$ is the current degree of $i$. Here we use
        $f(k_i)$ to constrain the nodes with higher degrees
        than the threshold $k_T$. We define
        \begin{equation}
        f(k_i)=\frac{1}{e^{\beta (k_i-k_T)}+1},
        \end{equation}
        where $\beta$ is a parameter to control the extent of
        the constraint.
        If $k_i>2$, randomly select two of its
        neighbors and establish a tie between them if they are not
        connected in the earlier stage.
        Repeat this action for
        \begin{equation}
        \frac{1}{2}c\sum_{i\in V}{k_i(k_i-1)}
        \end{equation}
        times, where $c$ is the speed of adding new ties.
    \item
        Action 2: Select a node $q$ with probability
        \begin{equation}
        p(q)=\frac{k_q+1}{\sum_{j\in V}{(k_j+1)}}.
        \end{equation}
        If $k_q>1$, select one of its neighbors randomly and remove
        the tie between them.
        Repeat this action for
        \begin{equation}
        \frac{1}{2}d\sum_{i\in V}{k_i}
        \end{equation}
        times, where $d$ is the speed of removing ties.
    \item
        Action 3: Randomly select a pair of nodes and add a tie between them if
        they are not connected originally. Repeat this action for
        $|V|r$ times, where $r$ is the speed of adding linkage randomly.
    \end{itemize}
\item
    Step 3: If the current averaged degree of the network reaches $\langle
    k\rangle_{max}$, stop evolving and return the network. Otherwise, increase the evolving time and then jump to Step 2.
\end{itemize}

{\bf Remark 1.} In the first step we choose to load a BA network
just because it is classical and simple. Many real-world networks
are found scale-free, including the social networks.

{\bf Remark 2.} In the Step 2 Action 1, the nodes with higher
degrees will be selected to increase the closeness of the network
among their friends. However, when the node's degree is higher than
the threshold $k_T$, the probability of getting selected will
decrease sharply. It responds to the situation that the nodes whose
degrees have exceeded the threshold, some of theirs online
friendships would be weak and it is hard for their neighbors to get
acquainted through them. This is the essential part of the model,
which is different from the others.

{\bf Remark 3.} In the Step 2 Action 2, the nodes with higher
degrees will be selected easily to loose a random acquaintance.
Because for the high-degree nodes, it is easy to ignore some
friendship for the constraints of time and cognition.

{\bf Remark 4.} In the Step 2 Action 3, a pair of nodes will be
randomly selected and connected. It responds to the phenomenon that
some friendships are established casually in online social sites.
For instance, some users may search other strange ones with common
interests or just accept some unknown invitations.

\begin{figure}[ht]
\centering \epsfig{file=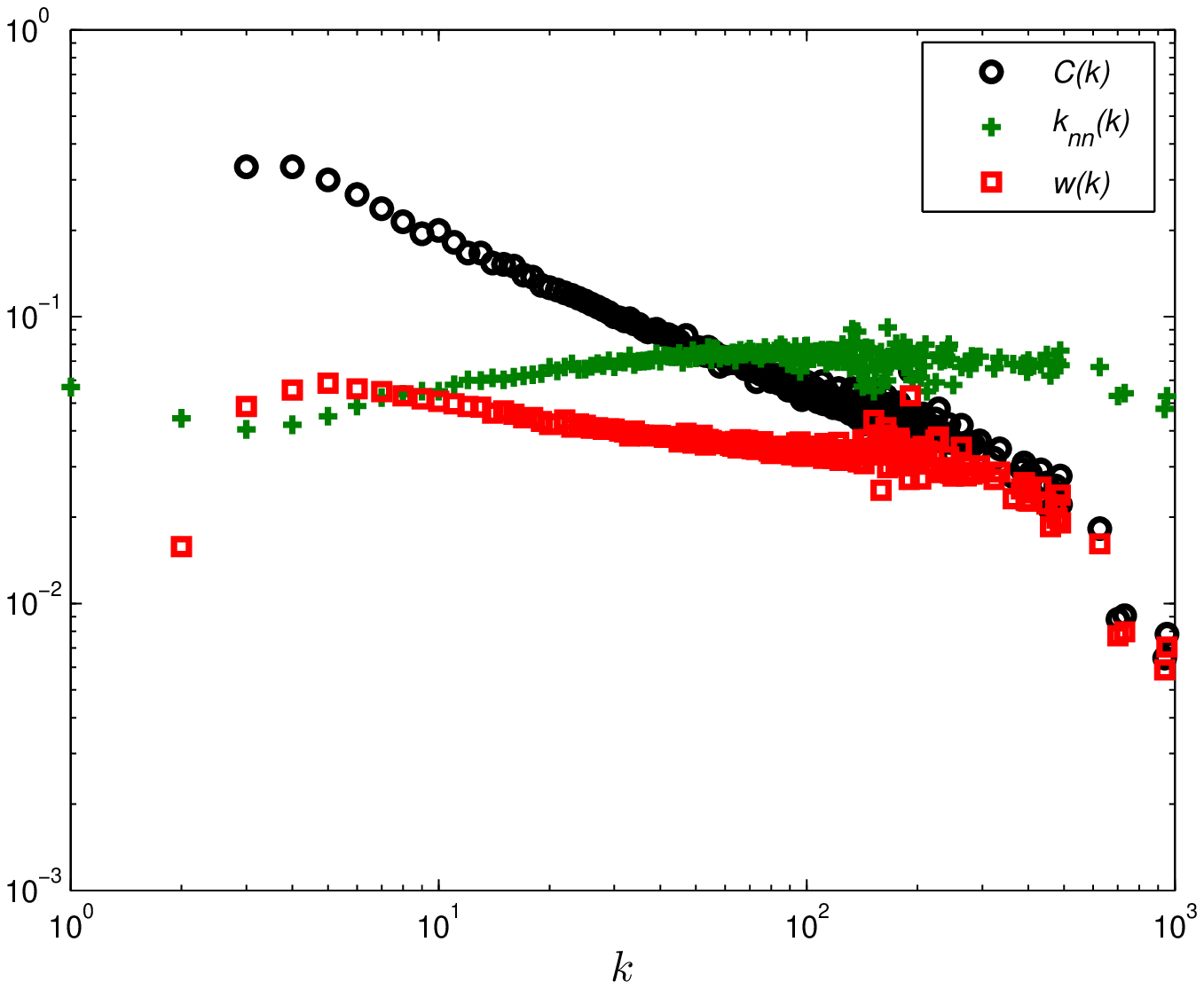,width=6cm}
\caption{\texttt{$BA-shift(20000,20,0.0005,0.0005,0.0001,8,200)$}}
\label{fig:model}
\end{figure}

In the following simulations, we denote the network generated by the
model as $BA-shift(|V|, \langle k \rangle_{max}, c, d, r, \beta,
k_T)$. The BA network we use contains 20000 nodes and 39973 ties
originally. As showed in Figure~\ref{fig:model}, it is easy to find
for all the local measures, including $C(k)$, $k_{nn}(k)$ and
$w(k)$, there exists an turning point near $k_T=200$. This result is
consistent with the real-world datasets. However, for the BA
network, the variations of $C(k)$ and $k_{nn}(k)$ keep decreasing
steadily with $k$, while $w(k)$ just increases without any
descending tendency. In fact, compared with other models, such as
JGN and BA, $BA-shift$ pays more attention to understanding how the
constraints of time and cognition affect the evolution of online
social networks. The aim of the model is to unveil the generation of
the threshold we find.

{\bf Summary} In this section ,we present a simple model to
interpret the generation of the upper limit. Compared to Dunbar's
number, the value of the limit in online social networks is greater.
We believe it will bring some impact and insight to the current
situation.

\section{Business Insights}
\label{sec:businessinsights}

\subsection{Online viral marketing}
\label{sec:viralmarketing}

\begin{figure}[ht]
\centering \epsfig{file=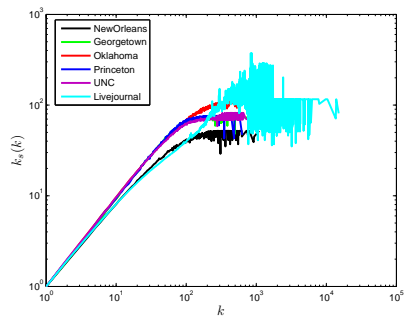,width=6cm} \caption{Variations of
$k_s(k)$ with $k$.} \label{fig:kshell}
\end{figure}

Thanks to the thorough growth of online social networks in the
recent decade, a new strategy for marketing has been deployed.
Nowadays we may often see some comments on a particular product from
our Facebook friends' wall, or advertisements may appear in the form
of tweets from the people we follow on Twitter. That's indeed an
instance of the online viral marketing, as product information
spreads from person to person directly(word-of-mouth) within the
networks and influences people's purchasing decisions. Viral
marketing in online social networks may be quite effective as people
may seriously consider friends' recommendations. However, questions
still remain on how to do it and where to start. Leskovec et al.
\cite{leskovec:dynamics} suggest that unlike epidemic spreading
models, high-degree nodes are not so influential in viral marketing
situation. This conclusion can be well supported by our
observations, since \lq\lq aggressive users\rq\rq~with thousands of
friends only interact with a small group of friends. We validate
this in a further step by importing the measurement of $k_s(k)$. As
shown in Figure~\ref{fig:kshell}, the averaged k-shell value stops
increasing and remains stable with slight fluctuations after the
threshold, meaning that the core effect is not obvious for
high-degree nodes. Just as illustrated in \cite{kshell-influential},
some high-degree users are wrapped by large amount of low-degree
users in the periphery, so that themselves are also positioned in
the periphery and play a trivial role in spreading product
information.

From what we have discussed in previous sections, we believe that
due to the loosely connected neighborhood and lacking of
interactions, the messages sent out by the \lq\lq aggressive
users\rq\rq~may be ignored easily. In contrast, \lq\lq rational
users\rq\rq~are tightly linked to each other, and these users'
online friends are largely covered by their offline friends.
Therefore they can be more trustful, and their messages would be
thought highly of by their friends. In consequence, their friends
may be induced by their purchasing suggestions and behaviors.

\subsection{Privacy management in online social networks}
\label{sec:trustandprivacy} In order to guarantee a more authentic
social network, most online social sites require users to provide
their authentic personal information when firstly register in the
sites. However, this can arouse the concerns for users' privacy
issues as users' profile information such as ZIP code, gender and
birthday may be stolen for improper use \cite{gross:information}.

At the same time, users have begun to recognize the necessities of
privacy protection in online social networks, especially the \lq\lq
rational users\rq\rq. Since these users have \lq\lq moved\rq\rq~
quite a number of offline friends to the social sites, they regard
the social sites as personal space to interact with their friends,
so that the interaction can be quite private, even secret. In view
of this, current social sites such as Facebook has already started
to provide the service of privacy settings. Users themselves can
determine whether to reveal their information only to friends or to
the public. In fact, friendships are regarded as binary in this
situation, that is to say, all the private settings are equally
effective to each friendship. However, as shown in our discussions,
users cannot treat each tie equally. \lq\lq rational users\rq\rq~
indeed have different attitudes toward their online friends, so they
may desire for a more detailed and flexible mechanism that enables
them to have different privacy settings for different groups of
friends.

However, things are different for \lq\lq aggressive users\rq\rq.
They do not care too much about privacy, and instead they are
willing to disclose their information to more users in order to gain
popularity. Another particular phenomenon should be noted is that
there exist \lq\lq spammers\rq\rq~in online social networks. \lq\lq
Spammers\rq\rq~ disguise as \lq\lq aggressive users\rq\rq, usually
with fake profiles of celebrities, to establish so many connections
with the intention to carry out identity theft \cite{haddadi:add},
which is a great threat to online users. To detect such \lq\lq
spammers\rq\rq, we can examine its interaction records because they
only add friends but do not interact at all.

\section{Conclusion}
\label{sec:conclusion}

Just as unveiled in social networks, there is still a magic upper
limit on users' number of friendships that they can effectively
maintain in online social networks. Through abundant experiments and
validations, we conclude that users with considerable circles of
friends within the magic number are \lq\lq rational users\rq\rq.
They mainly use online social networks on the purpose of maintaining
their old friendships. In contrast, \lq\lq aggressive
users\rq\rq~reach out of the magic number with the desire to make as
many new friends as possible. We also propose a new online social
network model to further explain users' online behaviors. We think
the findings of the new magic number and distinction of users are
helpful in viral marketing and privacy management issues in online
social networks.

\bibliographystyle{abbrv}
\bibliography{refs}  
\balancecolumns
\end{document}